\documentstyle[epsfig]{mn}

\title[XMM-EPIC Observation of MCG$-$6-30-15]{XMM-EPIC observation of
  MCG$-$6-30-15: Direct evidence for the extraction of energy from a
  spinning black hole?}

\author[J. Wilms et al.]{J\"orn~Wilms,$^1$
Christopher~S.~Reynolds,$^{2,3}$\thanks{Hubble Fellow while at Univ.~of~Colorado}
Mitchell~C.~Begelman,$^{3,4}$\newauthor
James~Reeves,$^5$
Silvano~Molendi,$^6$
R\"udiger Staubert,$^1$
Eckhard Kendziorra$^1$\\
$^1$Institut f\"ur Astronomie und
Astrophysik, Abt.\ Astronomie, Universit\"at T\"ubingen, Sand 1, D-72076
T\"ubingen, Germany\\ 
$^2$Dept.\ of Astronomy, University of Maryland, College Park, MD 20742\\
$^3$JILA, Campus Box 440, University of Colorado,
Boulder CO~80309\\
$^4$Dept.\ of Astrophysical and Planetary Sciences, University
of Colorado, Boulder, CO~80309\\
$^5$X-Ray Astronomy Group, Department of Physics and
  Astronomy, Leicester University, Leicester LE1 7RH, UK\\
$^6$Istituto di Fisica Cosmica, CNR, via Bassini 15,
  I-20133 Milano, Italy
}

\date{Accepted XXXX. Received 2001 August 22}
\pagerange{\pageref{firstpage}--\pageref{lastpage}}
\pubyear{2001}

\begin{document}
\label{firstpage}

\maketitle

\begin{abstract}
  We present {\it XMM-Newton} European Photon Imaging Camera (EPIC)
  observations of the bright Seyfert 1 galaxy MCG$-$6-30-15, focusing
  on the broad Fe~K$\alpha$ line at $\sim 6$\,keV and the associated
  reflection continuum, which is believed to originate from the inner
  accretion disk.  We find these reflection features to be
  \emph{extremely} broad and red-shifted, indicating its origin from
  the very most central regions of the accretion disk.  It seems
  likely that we have caught this source in the ``deep minimum'' state
  first observed by Iwasawa et al.\ (1996).  The implied central
  concentration of X-ray illumination is difficult to understand in
  any pure accretion disk model.  We suggest that we are witnessing
  the extraction and dissipation of rotational energy from a spinning
  black hole by magnetic fields connecting the black hole or plunging
  region to the disk.
\end{abstract}


\section{Introduction}
X-ray spectra of Seyfert~1 galaxies commonly show an iron K$\alpha$
emission line at $\sim$6\,keV.  The line is often extremely broad,
with a velocity width of $>70\,000\,\rm km\,s^{-1}$.  Furthermore, the
asymmetries in the line profiles are well explained by relativistic
effects.  Both of these facts suggest that the line is emitted from
the surface layers of the accretion disk within a few gravitational
radii $r_{\rm g}=GM/c^2$ of the black hole (BH) itself.  It is now
widely believed that this spectral feature is a relatively clean probe
of the immediate environments of supermassive BHs (see Fabian et al.,
2000, and references therein).

One of the best studied broad iron line sources is the nearby bright
Seyfert~1 galaxy MCG$-$6-30-15 ($z=0.008$).  This was the first active
galactic nucleus (AGN) for which a relativistic iron line disk profile
was measured.  In their {\it Advanced Satellite for Cosmology and
  Astrophysics} (ASCA) observations, Tanaka et al.
\shortcite{tanaka:95b} found that the iron line profile could be
explained by emission from an accretion disk around a non-rotating
(Schwarzschild) BH.  In a detailed reanalysis, however, Iwasawa et al.
\shortcite{iwasawa:96a} found a much broader line during a period of
low continuum X-ray flux, the so-called ``deep minimum state''.  The
line becomes so broad that disk models require emission from within
$6r_{\rm g}$, suggesting either a rotating (Kerr) BH, with its
marginally stable orbit at $r_{\rm ms}<6r_{\rm g}$ (Iwasawa et al.,
1996, Dabrowski et al., 1997), or iron fluorescence from material
spiraling into the BH at $ < r_{\rm ms}$ \cite{reynolds:97d}. Such a
broad line has later been confirmed, e.g., by Guainazzi et al.\ (1999)
and Lee et al.\ (1999).

Using the new generation of X-ray satellites such as
{\it XMM-Newton} with their improved collecting area and X-ray CCD
spectral resolution, the Fe line of Seyferts can be studied in greater
detail than was possible before.  In this {\it Letter}, we present
data from a 100\,ksec {\it XMM-Newton} observation of
MCG$-$6-30-15.

\section{Instrumentation and Data Reduction}
Our observation covered most of {\it XMM-Newton}'s orbit 108, on 2000
June 11/12, and was simultaneous with the {\it Rossi X-ray Timing
  Explorer} (RXTE). Here, we report on data from the European Photon
Imaging Cameras onboard {\it XMM-Newton} (EPIC, Str\"uder et al.,
2001, Turner et al., 2001).  To prevent photon pile-up, the EPIC-pn
camera was operated in its small window mode (using the medium thick
filter to prevent optical light contamination), and the EPIC MOS-1
camera was operated in its timing mode.  The MOS-2 camera was operated
in full-frame mode to study the field surrounding MCG$-$6-30-15.
Although we primarily rely on the pn-data here, the pile-up in the
MOS-2 data can be reduced to $<5\%$ when using single events only.
Therefore, we can use the MOS-2 data to check for the
instrument-independency of our results.

As of 2001~August, the official version of the {\it XMM-Newton}
Science Analysis Software (SAS) only contained a preliminary model for
the charge transfer efficiency (CTE) of the pn-chips. Therefore, we
used an internal SAS version implementing an improved model for the
EPIC-pn CTE (Haberl, priv.\ comm.).  To avoid remaining response
matrix uncertainties, we concentrate on the energy range from 0.5 to
11\,keV.  To produce EPIC-pn source and background spectra, we
collected events from two circles of radius 10~pixels on and off the
source for those times where the source count rate was below $14\,\rm
counts\,s^{-1}$.  This upper limit of the count rate is necessary to
avoid phases where the core of the MCG$-$6-30-15 point spread function
was slightly piled up.  The total resulting EPIC integration time was
54\,ksec. We checked the background lightcurve for periods of severely
increased background, but none were found.

For spectral modeling, response matrices appropriate for the improved
CTE model and appropriate for the source position on the EPIC chip
were used.  Furthermore, we corrected the exposure time in the spectra
for the $\sim$71\% livetime of the pn small window mode
\cite{kuster:99b} and perform simultaneous fits with single and double
events.  We estimate that the overall uncertainties in the spectral
calibration and flux calibration due to these procedures are at most a
few percent.

\begin{figure}
{\epsfxsize=\columnwidth\epsffile{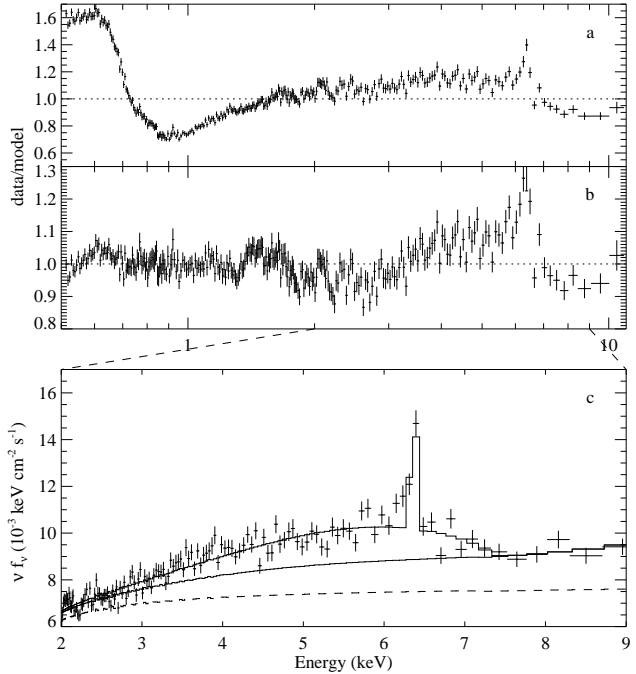}}
\caption{(a) Ratio between data and model from fitting a power-law
  to the 0.5--11\,keV data. (b) Ratio from fitting a power-law and the
  empirical warm absorber model (see text). (c) Deconvolved spectrum
  of the Fe K$\alpha$ band, showing the total {\tt laor} model and the
  continuum with and without (dashed) the reflection component for a
  model with reflection from an ionized disk. For clarity, the data
  have been rebinned and only the single event data points are
  shown.\label{fig:fits}}
\end{figure}

\section{The iron line profile and accretion disks fits}
Figure~\ref{fig:fits}a shows the ratio of the 0.5--11\,keV EPIC-pn
data to a spectral model consisting of a pure power-law (fitted in the
2--11\,keV band) modified by Galactic absorption ($N_{\rm H}=4.1\times
10^{20}\rm cm^{-2}$, using the XSPEC {\tt phabs} model with
cross-sections similar to those described by Wilms, Allen, \& McCray,
2000).  Below $\sim 2$\,keV, there is significant spectral complexity
which is due to a warm absorber (Nandra \& Pounds 1994, Reynolds \&
Fabian 1995, Lee et al.\ 2001), possibly superposed with a complex of
relativistically broadened soft X-ray recombination lines
\cite{branduardi:01a}.  At 3--7\,keV, a broad hump in the spectrum
suggests the identification with disk reflection signatures and,
especially, the broad iron K$\alpha$ line.

The putative broad iron line is more apparent once the soft X-ray
spectral complexity has been modeled.  This is not a trivial exercise
since the physical nature of this complexity is still a matter of
debate.  To start with, we consider a warm absorber scenario (this
will be generalized to include soft X-ray relativistic emission lines
below).  We assume that the continuum consists of a power-law modified
by Galactic absorption.  We also allow for the possibility of
intrinsic neutral absorption in MCG$-$6-30-15 (although the best
fitting column density is always consistent with zero).  By fitting
the high resolution X-ray spectrum from the Reflection Grating
Spectrograph (RGS), we construct an empirical warm absorber model
(based on Lee et al.\ 2001) with absorption edges of C\,{\sc v},
O\,{\sc vii}, O\,{\sc viii}, and Ne\,{\sc x} (at 392\,eV, 740\,eV,
870\,eV, and 1362\,eV, respectively, with best fitting optical depths
at threshold of $\tau=0.34$, $0.87$, $1.17$, and $0.10$), a Gaussian
absorption feature below the O\,{\sc vii} and O\,{\sc viii} edges to
crudely model blended resonance absorption lines, and four moderately
broad and weak Gaussian emission features at observed energies of
$0.885$\,keV, $0.935$\,keV, $0.995$\,keV, and $1.06$\,keV, which are
merely included to fit the RGS data; we do not attempt to make a
physical identification of these features.  After applying this
empirical model to the data, the soft X-ray residuals are reduced to
less than $10\%$, and the broad iron line feature is much more obvious
(Fig.~\ref{fig:fits}b).  The best fit photon index is $\Gamma=1.87\pm
0.01$ (goodness of fit $\chi^2/{\rm dof}=2625/1719$, all uncertainties
are quoted at the 90\% level for one interesting parameter,
$\Delta\chi^2=2.71$).

We now consider the nature of the hard spectral complexity.  While it
may be tempting to model the hump-like feature as pure broad iron line
emission, this is not a physically consistent model.  The equivalent
width of this feature (measured relative to the continuum at the iron
K$\alpha$ rest-frame energy) is $\sim 1$\,keV, and it would be
impossible to obtain such a broad iron line without also observing the
X-ray continuum photons that have been backscattered from the
accretion disk.  Consequently, we will not present pure broad iron
line models and jump, instead, to more physically motivated spectral
models which include both iron line fluorescence and the backscattered
``reflection'' continuum.  Within the framework of such models, the
3--7\,keV hump is a combination of iron line fluorescence and
backscattered continuum photons (with the Fe K edge responsible for
some fraction of the drop in flux at $\sim 7$\,keV).

We construct a spectral model appropriate for the case of reflection
from the Keplerian regions of an accretion disk around a near-extremal
Kerr black hole ($a=0.998$).  The inner edge of the X-ray reprocessing
region is taken to be the radius of marginal stability, $r_{\rm
  ms}=1.23r_{\rm g}$, and the outer edge is taken to be $r_{\rm
  out}=400r_{\rm g}$.  We fix the inclination of the inner disk to be
$i=30^\circ$ (Tanaka et al., 1995) and note that our results below
only slightly depend on the value of $i$ chosen. We assume that the
reflected emission has an emissity profile $\epsilon\propto
r^{-\beta}$ and leave $\beta$ as a free parameter.  The reflection
continuum is described using the {\tt pexriv} model of XSPEC and
normalized to the case where the accretion disk intercepts half of the
X-ray power-law continuum photons (i.e., $R=1$). Both, the iron line
profile and the reflection continuum are then relativistically smeared
(using the model of Laor, 1991, as the kernel).  Here, and throughout
the rest of the spectral modeling discussed in this paper, we also
include a narrow (unresolved) Gaussian line at 6.4\,keV with
equivalent width $W_{\rm K\alpha,N}$ to model the obvious narrow line
core.  This component presumably originates from distant material such
as the putative molecular torus of Seyfert unification schemes, and
has recently be seen in several other AGN (e.g., in NGC~5548, Yaqoob
et al., 2001, or in Mkn~205, Reeves et al., 2001).

Firstly, we consider the case of a fairly cold accretion disk
(rest-frame iron line energy of 6.4\,keV, disk ionization parameter
$\xi=1$, and $R=1$).  The resulting best fitting parameters are
$\Gamma=2.09\pm0.01$, $\beta=3.6\pm0.1$, $W_{\rm K\alpha,N}=38$\,eV,
and $W_{\rm K\alpha,B}=805$\,eV ($\chi^2/{\rm dof}=2383/1717$).
However, this model fails to explain the data as well as possessing an
unphysically large broad line equivalent width.  Significant residuals
(up to $\sim 20\%$) remain in the 3--11\,keV band.  The pattern of
residuals suggest that the model is failing to produce the red-wing of
the 3--7\,keV hump, and is underestimating the flux in the blueward
edge of this feature.  A hard tail above 8\,keV suggests that we are
also underestimating the strength of the high energy reflection
continuum.  If we allow the relative normalization of the reflection
continuum, $R$, to be a free parameter, the fit converges to a very
large value of $R$ ($R=7.6^{+1.8}_{-1.3}$, also $\Gamma=2.35\pm0.05$ and
$\beta=4.4\pm 0.2$, $\chi^2/{\rm dof}=2055/1716$). Despite the slightly
better $\chi^2$, large residuals around $\sim$7\,keV remain.

Ionization of the accretion disk surface is another way of increasing
the amount of reflection continuum.  To model this (with the relative
normalization of the reflection continuum fixed to $R=1$), we allow
the ionization parameter of the disk $\xi$, the temperature of the
disk $T$ and the energy of the iron line $E$ to be free parameters
($E$ is constrained to lie in the range of allowable Fe-$\rm K\alpha$
energies; 6.40\,keV to 6.97\,keV).  The resulting best fit has
$\Gamma=2.00\pm0.02$, $\beta=3.29^{+0.12}_{-0.11}$,
$E=6.40^{+0.05}_{-0}$\,keV, $\xi=1380^{+570}_{-190}$, $T>6.0\times
10^5$\,K, $W_{\rm K\alpha,N}=41$\,eV, and $W_{\rm K\alpha,B}=235$\,eV
($\chi^2/{\rm dof}=2051/1714$).  This leads to a significantly better
fit with the higher iron edge/line energies explaining the residual at
$\sim 7$\,keV.  However, significant residuals still exist in the
3--5\,keV range (suggesting that we are still under-predicting the
flux in the red-wing of the reflection feature) and above 8\,keV.

Detailed examination of this last spectral model shows that features
in the {\em soft} X-ray band below 1.5\,keV determine the best fitting
ionization parameter and temperature.  Consequently, such fits must be
viewed with suspicion --- there are a number of physical processes
which are not included in this reflection model but which may well
contribute to the soft X-ray spectrum (most notably, the recombination
line and edge emission from C, N, O, and Ne).  It would also not be
surprising if our empirical warm absorber model left small soft X-ray
residuals which may then be mistakenly fit by ionized reflection
models.

Firstly, we can consider the ionized reflection fits ignoring all data
below 1.5\,keV.  Refitting the ionized disk model (including the
empirical warm absorber) to the 1.5--11\,keV band results in a good
fit with no clear residuals.  The best fitting parameters are:
$\Gamma=1.94\pm 0.02$, $\beta=4.6\pm 0.3$, $E=6.95^{+0}_{-0.15}$\,keV,
$W_{\rm K\alpha,N}=51$\,eV, and $W_{\rm K\alpha,B}=582$\,eV
($\chi^2/{\rm dof}=1506/1312$).  $T$ and $\xi$ are unconstrained by
the data, while the relative normalization of the reflection continuum
is constrained to be $R<4.7$.  During the fitting $R$ is
anti-correlated with $W_{\rm K\alpha,B}$.  These quantities are
physically consistent with each other (i.e., the iron line has the
appropriate strength for that value of $R$ and $\xi$) when $R\sim
1.5$--$2$ (corresponding to $W_{\rm K\alpha,B}\sim 300$--$400$\,eV).

This spectral modelling suggests that the accretion disk may be
ionized.  Consequently, the K$\alpha$ line need not be intrinsically
narrow as we have assumed so far, but might be Compton broadened in
the hot skin of the accretion disk (note that Compton broadening alone
can not explain the large line width; Reynolds \& Wilms, 2000).  We
examine this possibility by relaxing our assumption of a narrow
intrinsic line and, instead, model the {\it intrinsic} iron line with
a Gaussian profile of width $\sigma$.  Using the 1.5--11\,keV data,
the best fitting value is $\sigma=0$ (with other parameters as above).
Even if we force the line to be very broad (e.g., $\sigma=$1\,keV, at
the expense of reducing the goodness of fit by $\Delta\chi^2=13$), the
data still require $\beta>3.9$.

The second approach for handling the soft X-ray spectrum is to
explicitly model the recombination K-lines of C, N, O and Ne.
Qualitatively, this model is the same as the mixed emission-line model
of Branduardi-Raymont et al.\ (2001) with the presence of additional
warm absorption. The best fitting parameters are $\Gamma=1.96\pm0.01$,
$\beta=4.70^{+0.25}_{-0.27}$, $E=6.97^{+0}_{-0.10}$\,keV, $W_{\rm
K\alpha,N}=53$\,eV, $\xi>130$ and $W_{\rm K\alpha,B}=546$\,eV
($\chi^2/{\rm dof}=1907/1704$).  Again, the relative reflection
normalization is poorly constrained by these data, but
self-consistency is achieved for $R\sim 1.5$--$2$ and $W_{\rm
K\alpha,B}\sim 300$--$400$\,eV.  We conclude that the parameterization
of the soft energy spectrum does not qualitatively change the results
for the broadened Fe K$\alpha$ line.

We note that the {\tt pexriv} reflection model employed here is also
incomplete due to the neglect of Compton smearing of the absorption
edges.  We do not expect this to seriously comprise our results since
the observed smearing is much greater than that expected from Compton
effects.  However, we will examine these effects using improved models
(Ballantyne, Iwasawa \& Fabian 2001; Nayakshin \& Kallman 2001) in a
future publication.  Preliminary investigations show our current
conclusions to be robust.  Finally, to check the dependence of these
result on possible remaining calibration uncertainties we used the
data from the MOS-2 camera.  The best fit parameters we find are
consistent to within the error bars, with slight deviations at the Si
edge that are a known calibration issue. We conclude that our results
are independent of any remaining calibration issues with the EPIC
instruments.

To summarize, our physically motivated best-fitting model consists of
reflection from an ionized accretion disk that emits H-like iron
K$\alpha$ fluorescence with a relative reflection fraction of
$R\approx 1.5$--$2$ and a broad iron line equivalent with of $W_{\rm
  K\alpha,B}\approx 300$--$400$\,eV. The 2--10\,keV flux of
$F_{2-10}=2.3\times 10^{-11}\,\rm erg\,s^{-1}\,cm^{-2}$ is comparable
with the ``deep minimum state'' found by Iwasawa et al.\ (1996,
$F_{2-10}=2.0\times 10^{-11}\,\rm erg\,s^{-1}\,cm^{-2}$). \emph{The
  most interesting feature of these spectral models is that a very
  steep emissivity profile $\beta=4.3$--$5.0$ for the
  iron-line/reflection features is required}.  We address the
implications of this result in the next section.

Finally, we note that if $r_{\rm in}$ is allowed to vary, the data
require $r_{\rm in}<2.1r_{\rm g}$ in order to model the red-wing of the
3--8\,keV hump.  We explicitly note that the data cannot be adequately
described by any reflection model with an inner radius of $r_{\rm
  in}=6r_{\rm g}$, the radius of marginal stability around a
Schwarzschild black hole.  Thus, models in which the X-ray reflection
occurs in the Keplerian part of accretion disks around a non-rotating
black hole cannot explain these data.  In principle, there can be
X-ray reflection inside of the radius of marginal stability (Reynolds
\& Begelman 1997). However, as discussed below, the extremely steep
emissivity index $\beta$ required by our data is very hard to
understand in the context of Schwarzschild geometry.

\section{Discussion}\label{sec:discuss}
Very steep emission profiles are required in all of our good fits to
the Fe K$\alpha$ line and reflection continuum.  Here, we explore the
implications of this result, assuming that the reflected flux
(including the iron line) from a local patch of the disk is
proportional to the X-ray continuum flux (at the iron K-shell edge)
irradiating that patch.  We will assume that a fixed fraction $f$ of
the energy released locally in the body of the accretion disk is
transported into an accretion disk corona and then radiated in the
X-ray band.  Studies of such coronae suggest that $f\sim 1$ in order
to produce the observed continua (Haardt \& Maraschi 1993, Dove et al.
1997).  Thus, while our assumption may not be true in detail, it must
be approximately true across a wide range of disk radii.

\subsection{Accretion disk models}
Can standard BH accretion disks models (Novikov \& Thorne 1973,
Riffert \& Herold 1995) explain the observed emissivity profiles?
With the assumptions described above, this question reduces to an
examination of the radial distribution of energy dissipation in such
disk models.  For a disk around an $a=0.998$ BH, the flux emitted from
the disk per unit proper area of the disk, ${\cal E}(r)$, peaks at
$r\sim 1.6r_{\rm g}$ and then gradually steepens to approach ${\cal
E}(r)\propto r^{-3}$. At no point does ${\cal E}(r)$ become as steep
as $r^{-4.5}$, as required by our iron line observations.  This is
true for any assumed BH spin.  Also, as noted above, the emissivity
profile is steeper than accretion disk models even if one accounts for
intrinsic broadening of the iron line via Comptonization.

We note that there are two complications that may be relevant to real
disks, even though they go beyond the realm of standard disk models.
Firstly, magnetic fields might couple material within $r=r_{\rm ms}$
to the rest of the disk, thereby permitting continued energy
extraction from this material (Krolik 1999; Agol \& Krolik 2000,
hereafter AK00).  Non-relativistic accretion disk simulations
employing pseudo-Newtonian potentials suggest that the stresses and
presumably the dissipation remain fairly flat within this region
(Hawley \& Krolik 2001; Armitage, Reynolds \& Chiang 2001).  Hence, it
would seem that the required emissivities are difficult to achieve
through such effects for disks around Schwarzschild holes.  Secondly,
X-rays produced away from the equatorial plane might be
gravitationally focussed into the central regions of the disk.
However, using the method of Petrucci \& Henri (1997), we find that
such effects cannot produce the observed emissivity profile unless the
X-ray source is already situated in the very central regions of the
disk.

We conclude that the disk has to be irradiated in a more centrally
concentrated manner than predicted by current pure-accretion disk
models.  In other words, we require some additional X-ray source that
is both powerful and very centrally concentrated.  There is one
obvious candidate: {\it X-rays that are associated with the magnetic
extraction of BH spin energy.}

\subsection{Magnetic extraction of BH spin energy}

Here, we explore the suggestion that the central X-ray source is
associated with the extraction of BH spin energy, focusing on
scenarios in which the BH spin is extracted via magnetic fields that
pierce the (stretched) event horizon \cite{blandford:77a}.  An
alternative class of models in which the magnetic fields do not pierce
the horizon will be addressed below. In order to produce the required
steep irradiation profile, the X-ray production must occur close to
the disk itself with a production rate that is a steeply declining
function of radius.  Thus, we are led to consider models in which the
magnetic field lines connect the rotating event horizon to the
accretion disk and/or disk corona.  For definiteness, we will consider
the canonical near-extremal Kerr BH.

In models in which the magnetic field lines are strongly coupled to
the body of the accretion disk, the magnetic field transmits a
retarding couple to the BH if the BH rotates faster than the
magnetically connected region of the disk where the extracted
rotational energy will be deposited. In order to power the observed
X-ray source, the field strength close to the BH needs to be $B\sim
10^4\,{\rm G}\,(M/10^7\,M_\odot)^{-1}$ \cite{blandford:77a}.  This is not
an impossibly high field; the corresponding magnetic pressure is still
substantially below the ram pressure of the accretion flow within
$r_{\rm ms}$ and so can be confined to the BH region.

If a ring within the disk of width $\delta r$ is connected to the event
horizon with magnetic flux $\delta \Psi$, then the power dumped into
the ring is (setting $GM_{\rm BH}=c=1$)
\begin{equation}
\delta P=\frac{(\delta \Psi)^2 \Omega_{\rm D}(\Omega_{\rm H}-\Omega_{\rm
D})}{4\pi^2 {\delta r} (-dZ_{\rm BH}/dr)},
\end{equation}
where $\Omega_{\rm D}(r)=(r^{3/2}+a)^{-1}$ is the angular velocity of
the accretion disk, and the angular velocity of the event horizon is
$\Omega_{\rm H}=0.479$ (Li 2000).  The BH resistance, $Z_{\rm BH}$, is
a function of disk radius, $r$, defined by a map from the BH horizon
to the accretion disk along the magnetic field lines.  Viscous forces
then transport this energy {\em outwards} by some distance before it
is dissipated and radiated.  AK01 and Li (2000) examine this process
in detail and compute the radial dependence of the energy dissipation
when the magnetic field connects to the accretion disk at one
particular radius.  We can use the formulae of AK01 (also see Li 2000)
as the Green's function for computing the more general case.

It is beyond the scope of this letter to address the detailed
distribution of the hole-threading magnetic field across the disk
surface, $B(r)$, or the nature of $Z_{\rm BH}(r)$.  If all of the
extracted spin energy is dumped into a ring at $r=r_{\rm ms}$, the
emissivity is very steep ($\beta>6$) within $r<1.8r_{\rm g}$ and
gradually flattens to ${\cal E}\propto r^{-3.5}$ at large radius
\cite{li:01a}.  More generally, the spin energy will be deposited into
a range of radii, flattening this profile.  We will assume $dZ_{\rm
BH}/dr\propto r^{-\nu}$ and $d\Psi/dr\propto r^{-\mu}$ ($\mu=2$
corresponds to a dipole field for $r\gg 1$).  For $\nu=1$ and $\mu=2$,
${\cal E}\propto r^{-3}$ at $r=2r_{\rm g}$ and is flatter inside that
radius.  To produce emission profiles as extreme as those required by
our data (i.e., $\beta\sim 4.5$ at $r\sim 1.5r_{\rm g}$), field
configurations as concentrated as $\mu\sim 3$--4 (with some dependence
on $\nu$).  Such conditions could be achieved, for example, if the
field were ``pinned'' onto the black hole by the ram pressure of the
accretion flow.

We note that the models addressed by AK00 do not employ the
Blandford-Znajek effect but, instead, magnetically torque the disk via
coupling to {\it matter} deep within the plunging region.  For
parameters relevant to our discussion, the extra energy source is
provided by the black hole spin via the Penrose effect occuring within
the radius of marginal stability (but outside of the stretched
horizon).  In this case, all of the extra torque is provided at the
inner edge of the disk (rather than across a range of radii as
addressed above) and so a steep emissivity profile is a natural
outcome.

Finally, we note that the large self-consistent value of $R\sim 1.5-2$
may have its origin in General Relativity.  Some fraction $f_{\rm
ret}<0.5$ of the upwardly directed disk emission will be bent by the
strong gravity and strike the disk again, (``returning radiation'',
Cunningham 1975, Speith, Riffert \& Ruder 1995, AK00) further
enhancing $W_{\rm K\alpha,B}$.  This can enhance the relative amount
of reflection by up to a factor of 2.  Further computations are
required to assess the effect of returning radiation on the emissivity
profile in a self-consistent manner.

\section{Conclusions}

We have presented {\it XMM-Newton}-EPIC observations of MCG$-$6-30-15
containing a spectral feature that is best described as an extremely
broad and redshifted X-ray reflection feature.  Both on the basis of
the extreme spectrum and the source flux, it seems likely that we have
caught MCG$-$6-30-15 in the peculiar ``deep minumum'' state first
noted by {\it ASCA} (Iwasawa et al. 1996).  The extreme nature of the
line profile leads us to conclude that it originates from the
centralmost parts of the accretion disk.  Standard accretion disk
models cannot produce an emissivity profile which is centrally
concentrated enough to produce the observed feature.  It also seems
unlikely that gravitational focussing of the continuum X-rays or
magnetic coupling of the plunging region to the rest of the disk can
ameliorate this conclusion.  Therefore, we suggest that, during the
deep minimum state, X-rays associated with the magnetic extraction of
the spin energy of the black hole are dominating the emission,
producing a sufficiently compact source to explain our observations.
Our results thus confirm the suggestion of Iwasawa et al.\ (1996) and
Dabrowski et al.\ (1997) that MCG$-$6-30-15 possesses a rapidly
rotating black hole.

In a forthcoming paper, we will present a more detailed analysis of our
observation, including simultaneous fits with the RXTE PCA, a study of the
time dependence of the reflection features and the X-ray continuum, and
fits using a larger and more detailed set of continuum and reflection
models.

\section*{acknowledgments}

We thank A.C.~Fabian, F.W.~Haberl, J.H.~Krolik, and M.~Kuster for useful
conversations, and the referee, Chris Done, for her very prompt reply
and useful comments.  We acknowledge support from DLR grant 50~OX~0002
(JW, RS, EK), Hubble Fellowship grant HF-01113.01-98A (CSR), and NSF
grant AST~98-76887 (CSR, MCB). This work is based on observations
obtained with {\it XMM-Newton}, an ESA science mission with
instruments and contributions directly funded by ESA Member States and
the USA (NASA).

\label{lastpage}

\end{document}